\documentclass[letterpaper]{jpconf}
\usepackage{graphicx}
\begin{document}
\title{High energy cosmic rays }

\author{Graciela B. Gelmini}

\address{Department of Physics and Astronomy, UCLA,\\
475 Portola Plaza, Los Angeles, CA 90095, USA}

\ead{gelmini@physics.ucla.edu}

\begin{abstract}
I review here  some of  the physics we are learning and expect to learn in the near future through the observation of cosmic rays. The study of cosmic rays involves a combination of data from accelerators, ground arrays, atmospheric fluorescence detectors and balloon and satellite experiments. I will discuss the data of the Pierre Auger Observatory, PAMELA, ATIC and FST among other experiments\footnote[1]{Plenary talk at {\it ``Discrete `08"},
11-16 December, 2008, Valencia, Spain}.
\end{abstract}

\section{Introduction}
A multi-messenger approach is necessary to  study the most energetic  particles in nature. We expect the sources to produce cosmic rays (i.e. charged particles), photons and neutrinos.   Their detection requires multiple techniques. Cosmic rays are detected in ground arrays and atmospheric fluorescence telescopes (i.e.  the Pierre Auger Observatory), detectors in balloons (e.g. HEAT, ACT) and satellites (e.g. PAMELA, Fermi Space Telescope) and  the space station (AMS).  High energy gamma rays are studied with  air cherenkov telescopes (ACT,  such as HESS, VERITAS, CANGAROO, MAGIC), detectors in satellites (e.g. INTEGRAL and Fermi), ground arrays and atmospheric fluorescence telescopes (Auger). High energy neutrinos  require under ice or water kilometer cube detectors, such as IceCube in Antarctica or the KM3NeT in the Mediterranean See,  and  radio telescopes in balloons, such as ANITA . I will concentrate here on cosmic rays and will mention photons (see the talk of Francis Halzen  on neutrinos in this conference). 

Cosmic rays have a non-thermal spectrum well approximated by a power law close to $\sim E^{-3}$ for 10 orders of magnitude in energy, from $10^{10}$ eV to above $10^{20}$ eV,  and 30 orders of magnitude in the flux. The flux goes from 1 particle/(m$^2$ sec)  at 10$^{11}$ eV, to
 1 particle/(m$^2$yr) at 10$^{16}$ eV,  to  1 particle/(km$^2$yr) at 10$^{19}$ eV (see Fig.~1). The non-thermal spectrum of cosmic rays indicates that their acceleration results from stochastic processes in the presence of magnetic fields, as first proposed by Fermi. The diffusive acceleration of charged particles  in the shock waves of supernova explosions may give origin to most cosmic rays. But the highest energies of cosmic rays are so extreme that it is not understood if a similar mechanism can accelerate these particles even in the most extreme sources, which may or may not be large enough  and have sufficiently large magnetic fields.
 
 Each energy range of cosmic rays addresses different physics. From 10$^8$ eV to  10$^{10}$ eV the modulation of the incoming charged particles by the solar wind can be studied. The expanding plasma generated by the Sun decelerates and partially sweeps-off the lower energy galactic cosmic rays from the vicinity of the Earth. This effect depends on the level of solar activity, which has an 11 year cycle. From 10$^{10}$ eV to 10$^{17}$ or 10$^{18}$ eV, cosmic rays allow us to study the galactic sources that produce them and the  propagation of charged particles within the galaxy. 
Above this energy range and up to at least a few 10$^{20}$ eV the cosmic rays must be of extragalactic origin although the characteristics of the transition is still not understood. Thus the so called ultra-high energy cosmic rays (UHECR), those above 10$^{18}$ eV, allow us to study their extragalactic sources and the propagation of charged particles in the extragalactic medium. 

\begin{figure}[h]
\includegraphics[width=20pc]{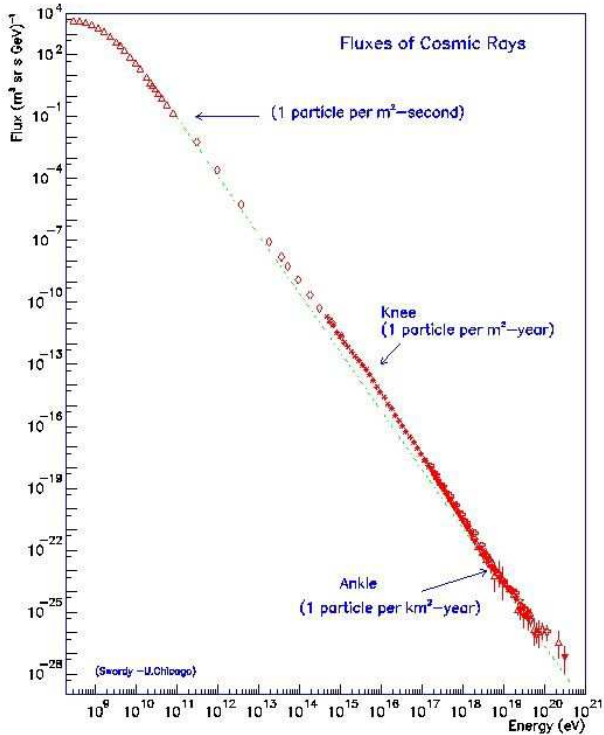}\hspace{2pc}%
\begin{minipage}[b]{13pc}
\caption{\label{fig1} Cosmic ray spectrum $\sim E^{-3}$ over 10
orders of magnitude from $10^{10}$ eV to above $10^{20}$ eV(from Ref.~\cite{CR})}
\end{minipage}
\end{figure}

\section{The 10$^8$ to 10$^{12}$ eV energy range}

The PAMELA satellite (Payload for Antimatter Exploration and Light-nuclei Astrophysics), consists of a magnetic spectrometer in orbit, launched in June 2006, and has produced its first data releases in 2008. Of particular interest has been the positron fraction $e^+/ (e^+ + e^-)$ data the PAMELA collaboration announced first in July 2008~\cite{Adriani:2008zr} (see Fig.~1), which shows an excess in the 10 -100 GeV range with respect to what is expected from secondary cosmic rays, i.e.  a fraction diminishing as the energy increases (secondary electrons and some positrons are produced in the interactions of cosmic ray protons and heavier nuclei). This result confirms what was earlier called the``HEAT excess"~\cite{HEAT}, also observed by AMS-01~\cite{AMS}. Fig.~1 shows the PAMELA data (in red) and the data of several prior experiments (in black). The large dispersion in the data seen at energies below a few GeV is expected, because the different experiments were done in different moments of the solar cycle. Above 10 GeV the solar modulation is not important and this is the range where the excess is significant. As seen in Fig.~1, 
the positron fraction is about 0.05 at 10 GeV and grows steadily with energy to reach 0.1-0.2 at  100 GeV.

 Soon after the publication of the PAMELA positron data, in November 2008, the ATIC collaboration announced a 6$\sigma$ excess in the $(e^+ + e^-)$ flux in the range 300 to 800 GeV~\cite{ATIC}, with respect to what is expected from secondary cosmic rays. If the PAMELA positron fraction excess is due to an excess of positrons (as opposed to a defect of electrons, which are much more abundant than positrons in secondary cosmic rays) it may mean that positrons and electrons are been created by an unknown  source. The excess electron flux in the PAMELA energy range would not be noticed until it  becomes  comparable to,  and then larger than, the secondary electron flux as the energy increases. If this is so, ATIC may have measured the continuation to higher energies of the excess found by PAMELA.
 
 \vspace{0.5cm}
 
\begin{figure}[h]
\begin{minipage}{19pc}
\vspace{-1.5cm}
\includegraphics[width=20pc]{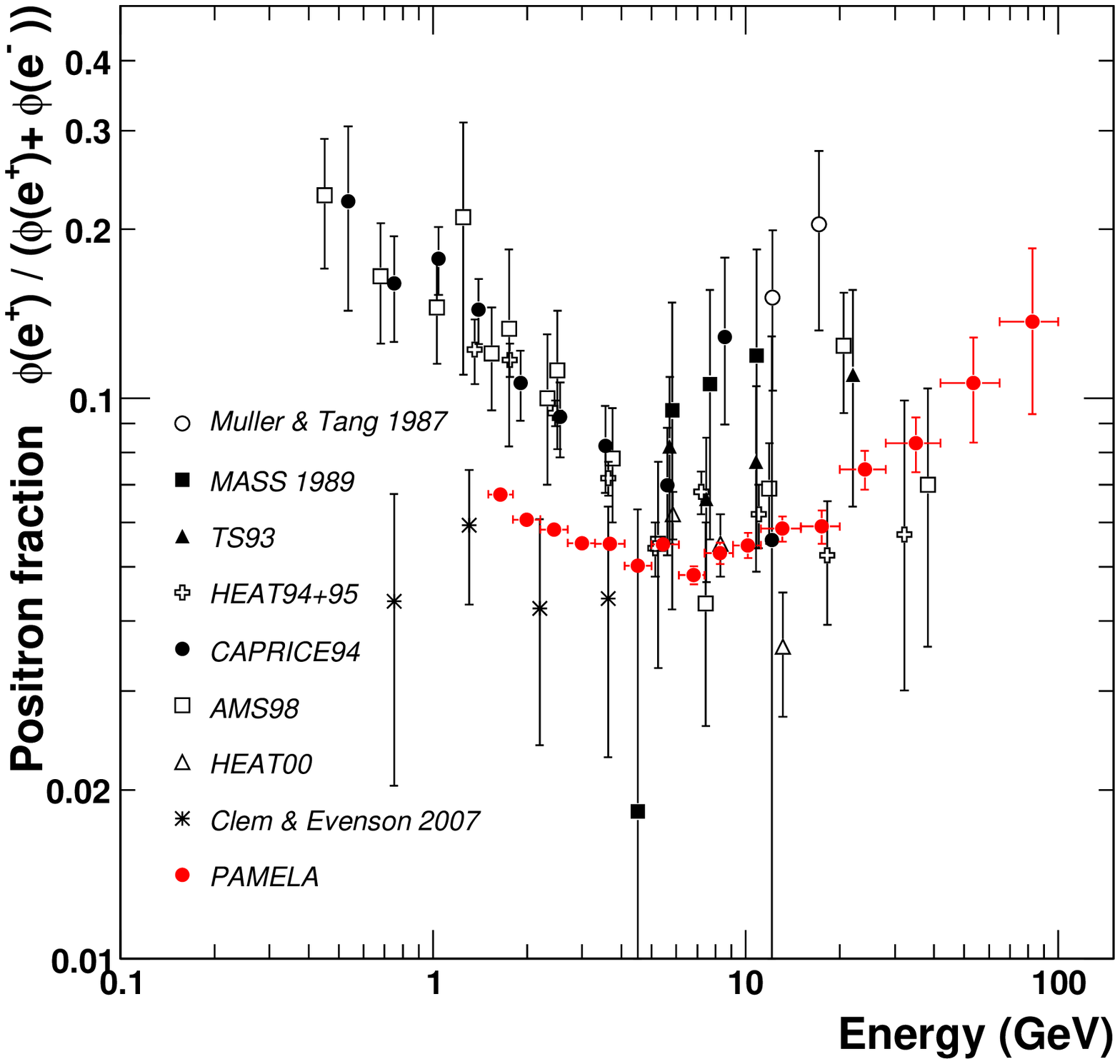}
\caption{\label{fig2} Positron fraction as function of energy measured by PAMELA~\cite{Adriani:2008zr} (in red) and several other earlier experiments.}
\end{minipage}\hspace{2pc}%
\begin{minipage}{18pc}
\vspace{-0.7cm}
\includegraphics[width=17pc]{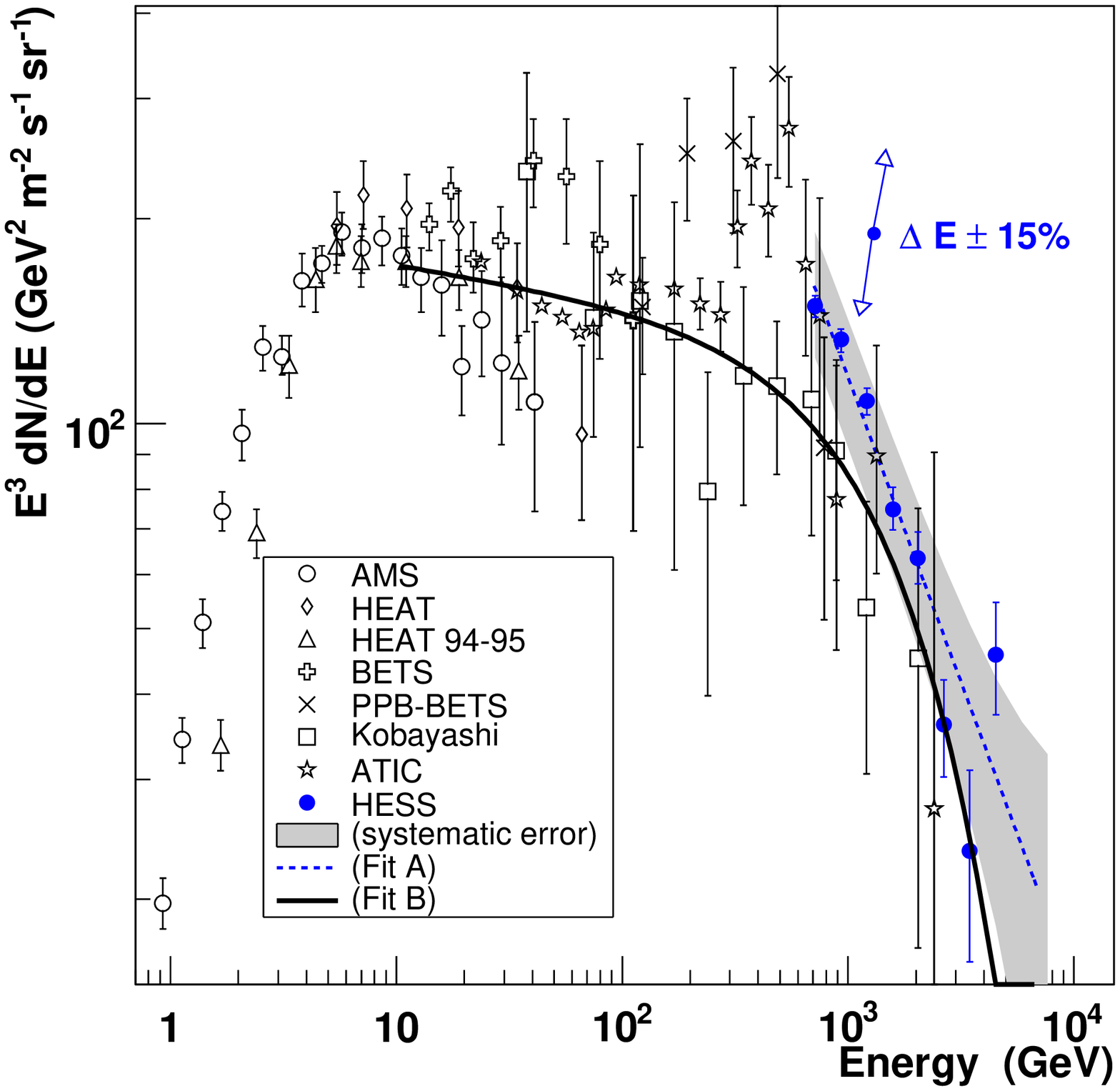}
\caption{\label{fig3} Energy spectrum of $(e^+ + e^-)$ multiplied by $E^3$ measured by HESS~\cite{HESS}, ATIC~\cite{ATIC} and other experiments.  The double arrow  shaded area indicate the approximate range of the systematic error in the HESS spctrum (from Ref.~\cite{HESS})}
\end{minipage} 
\end{figure}

 ATIC (the Advanced Thin Ionization Calorimeter instrument) is a balloon born calorimeter launched from the McMurdo base in Antarctica, which has flown several times. The data presented are from the 2000-2001 and 2002-2003 flights. A few days after the announcement of ATIC, the HESS collaboration published its $(e^+ + e^-)$ flux measurement  above 600 GeV, which  is steeply falling with increasing energy~\cite{HESS}.
 
  Fig.~2 (from Ref.~\cite{HESS}) shows the data of HESS, ATIC and other earlier experiments. The  $(e^+ + e^-)$ flux is multiplied by $E^3$. The double arrow shows the possible shift of the HESS spectrum due to  $\simeq 15$\% energy scale uncertainty  and the shaded area indicates the approximate range of the systematic error in the HESS spectrum due to other reasons. 
  
   The HESS data show that,  if the ATIC data are correct, there is a sharp cutoff of any excess above about  800 GeV, which is compatible with the sharp cutoff expected from dark matter particle annihilation at the parent particle mass. Since their appearance, at the moment of writing these notes the data of PAMELA and ATIC have generated more than 50 papers, at the rate of several per week, providing different explanations for their excesses.

The main two explanations are the existence of an astrophysical source nearby, or dark matter particles annihilating nearby, within a volume around Earth much smaller than the volume of the galaxy.  Electron and positrons rapidly loose energy through synchrotron and inverse Compton processes, thus at the energies observed they must come from less than 1 kpc away (for comparison, the distance between the Sun and the center of the galaxy is about 8 kpc). Known possible astrophysical sources are pulsars, of which there are some, such as Geminga and Vela, within a few 100 pc  from Earth~\cite{pulsars}. It is not known the mechanism through which they would produce such energetic electrons and positrons, thus the spectrum they could produce is unknown too. The maximum energy cutoff of the spectrum observed would in this case correspond to the maximum energy of emission of positrons and electrons. 

In the case of annihilating dark matter particles, the cutoff in energy would correspond to the mass of the parent particle. The cutoff could be very sharp if the particles annihilate directly into $e^+ e^-$ pairs, and this would be a clear indication of  the dark matter origin of the signal~\cite{Hall-Hooper} (see Fig.~4).  If instead $e^+ e^-$ are produced after a chain of decays of annihilation products, the cutoff would be less sharp, similar to what would be expected from astrophysical sources. Fig.~4 (taken from Ref.~\cite{Hall-Hooper}) shows  three fits to the ATIC data, assuming an astrophysical source, a dark matter particle annihilating directly to $e^+ e^-$ and another producing $e^+ e^-$ as secondaries of the annihilation products.

  In order to account for the PAMELA and/or ATIC data, the annihilation rate of most dark matter candidates needs to be enhanced by a factor, called ``boost factor" $B$, which may reach $B \simeq 10^{4}$ in the case of many usual WIMP dark matter candidates~\cite{Bergstrom:2008gr} and may be much smaller, close to $B \simeq 10$ (see Fig.~4),  for some candidates that decay preferentially into electrons and positrons (the ATIC collaboration finds that a Kaluza-Klein particle decaying mostly into electrons and positrons with mass 620 GeV fits well their electron-positron data~\cite{ATIC}).
\begin{figure}[h]
\includegraphics[width=20pc]{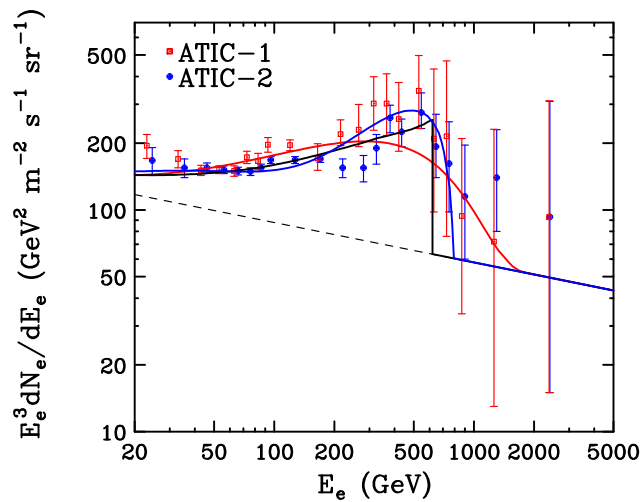}\hspace{2pc}%
\begin{minipage}[b]{13pc}\caption{\label{fig4} Examples of fits to the ATIC data assuming a pulsar nearby (red), the annihilation of  800 GeV dark matter particles producing $e^+ e^-$ as secondaries with $B \simeq 10^4$ (blue) and the annihilation of 620 GeV Kaluza-Klein dark matter particle decaying directly into $e^+ e^-$
with $B\simeq 10$ (black). From Ref.~\cite{Hall-Hooper}}
\end{minipage}
\end{figure}

Several ways of enhancing the local annihilation rate have been proposed:  there could be a  dense dark matter clamp nearby (however, the probability found in structure formation simulations for this to happen is about 10$^{-4}$~\cite{Vogelsberger}), which would enhance the rate, which depends on the square of the dark matter density,  or the annihilation cross section itself could be enhanced. For example  a ``Sommerfeld resonance" results from the almost formation of a bound state by the annihilating particles at very small relative velocities. This effect would enhance the annihilation cross section in the dark halo (but not in the early Universe, where the particles move at larger velocities)~\cite{Hisano:2004ds}. 

Whatever produces the positron excess seen by PAMELA should not produce an excess of antiprotons, 
since the antiproton to proton ratio $\bar p/p$ was also measured by PAMELA~\cite{Adriani:2008zq} and is compatible with what is expected from secondary cosmic rays. This could be partially explained because there are many more protons than electrons  in the secondary cosmic rays (PAMELA requires a  10$^{-5}$  discrimination  of protons to measure positrons) and because $p$ and $\bar p$ reach us from a larger volume than $e^+$ and $e^-$, from a large fraction of the total volume of the galaxy.

The Fermi Space Telescope (ex GLAST-Gamma-ray Large Area Space telescope) can also detect $e^+$ and $e^-$ (it does discriminate between the two), besides photons.  Fermi was launched in June 2008 and is providing $\gamma$-ray spectroscopic data of unprecedented quality. It has two detectors (the GLAST Burst Monitor, GMB, and the Large Area Telescope, LAT) which detect photons between 10 keV and 300 GeV, and observe the whole sky every 3 hours. Photons of even higher energy will be detected by Air Cherenkov Telescopes (ACTs), Cangaroo III in Australia, HESS in Namibia, MAGIC in Las Palmas and Veritas in the US (see the talk of Manel Martinez in this conference). These can also detect $e^+ +e^-$, as we already mentioned in the case of HESS.

 The Universe is totally transparent to photons below  100 GeV. At higher energies photons interact with infra-red and optical backgrounds but still at energies below 10's of TeV arrive to us from cosmological distances (at higher energies they are absorbed by the cosmic microwave background radiation). Photons can again reach earth from cosmological distance at energies above $10^{10}$ GeV, the range of UHECR. 
 
 The photons observed by Fermi reveal, therefore, the spacial distribution of their sources. In particular, if
 Fermi observes the annihilation of dark matter particles into photons, it is expected to detect not only the center of our galaxy but also a large amount of dark matter clumps. As mentioned above, since the annihilation rate depends on the square of the density, high density regions boost the rate. A large amount of dark matter clumps are expected to remain within the dark halo of our galaxy. Haloes grow hierarchically, incorporating lumps and tidal streams from earlier phases of structure formation. The best
 simulations of structure formation at present (the Via Lactea II~\cite{Kuhlen:2008aw} and the Aquarius Project~\cite{Vogelsberger}) show that 10's of subhaloes could be discovered by Fermi. However, subhaloes are more effectively destroyed by tidal effects near the center of the galaxy, thus most of them are far from the Sun. The chance that a random point close to the Sun  is lying within a clump is smaller than 10$^{-4}$~\cite{Vogelsberger}.  Dwarf galaxies and other objects which Fermi might observe as a whole  through dark matter annihilation into photons will have the annihilation rate increased  by a boost factor $B$, may be as large of 100, due to the existence of substructure within them (with respect to the rate they would have without dark matter lumps). But the concept of a boost factor cannot be applied to the possible enhancement of a signal in electron and positrons as seen by PAMELA or ATIC, since at most only one or very few dark matter clump could be present within the volume these particles come from.
 
The Fermi Space Telescope can measure the $(e^+ + e^-)$ flux in the range explored by ATIC, possibly to 1TeV, with better accuracy and its results are expected to be made public soon, in the April 2009 Meeting of the American Physical Society (which will take place in the first week of May 2009).

\section{The 10$^{13}$ to 10$^{18}$ eV energy range}

\begin{figure}[h]
\includegraphics[width=37pc]{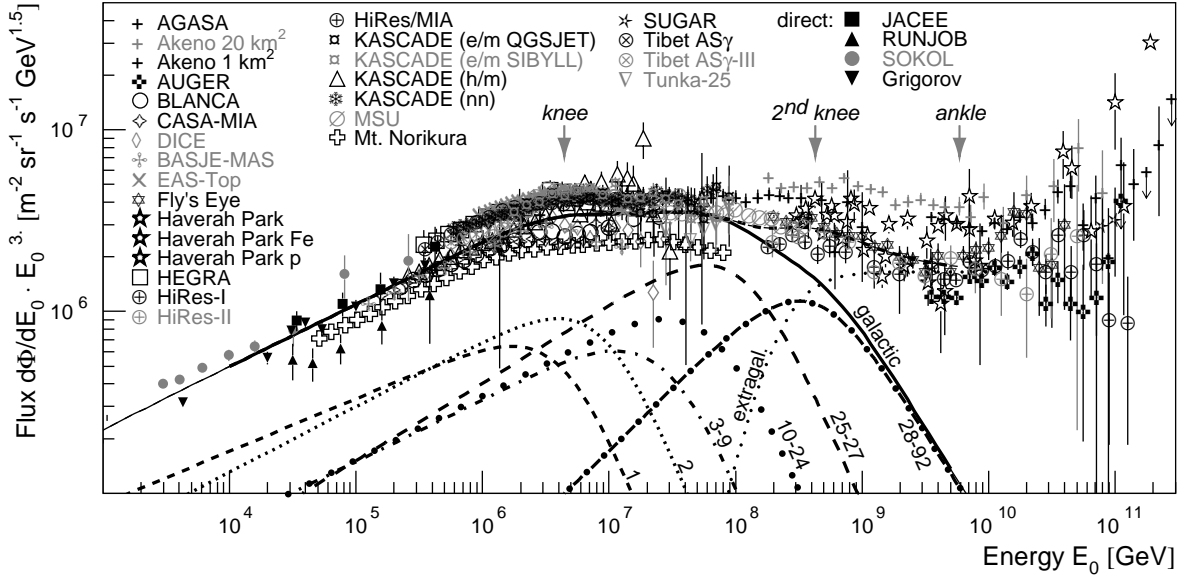}
\caption{\label{fig5} All-particle energy spectrum of cosmic rays, the flux is multiplied
	   by $E^3$ (from Ref.~\cite{Horandel:2006jd}). The lines represent
	   spectra for elemental groups $Z$ as
	   indicated according to a particular model. The sum
	   of all elements (galactic) and a presumably extragalactic component
	   are shown as well (see Ref.~\cite{Horandel:2006jd} for details).}
\end{figure} 
In Fig.~1 two features of the cosmic ray spectrum are indicated: the ``knee" above 10$^{15}$ eV and the
``ankle"  above 10$^{18}$ eV. These are shown also in Fig.~5 (taken from Ref.~\cite{Horandel:2006jd}) in which the flux has been multiplied by a factor $E^3$. At the knee there is a slight steepening of the spectrum, the spectral index changes from $- 2.7$ at low energies to about {-3.1} at higher energies. A ``second knee" feature is sometimes mentioned, at about 4$\times 10^{17}$ eV, where the spectrum seems to become again steeper at higher energies  with a spectral index approximately $-3.3$ (see Fig.~5). At the ankle the spectrum seems to flatten again to a spectral index of about $-2.7$. The ``ankle" is sometimes seen as just part of a  ``dip" in the spectrum between 10$^{18}$ and 10$^{19}$  eV.

At energies below 10$^{14}$ eV the elemental composition of cosmic rays is well known because cosmic ray  particles are detected directly.  The composition and flux of cosmic rays at these energies are in agreement with their production in galactic supernova explosions~\cite{Gaisser:2006sf}. Direct measurements above the atmosphere  to energies slightly above 10$^{14}$ eV are performed by various instruments, such as ATIC, CREAM, BESS and TRACER. At higher energies, the flux of cosmic rays becomes smaller than several particles /(m$^2$ day) so their detection with instruments carried in balloons or spacecrafts becomes impractical  and cosmic rays are detected from the ground through the shower of secondary particles they produced in the atmosphere. These are called Extensive Air Showers (EAS).  Arrays of detectors  covering a large area, usually from a fraction to many km$^2$, operate for years to obtain enough statistics. Because of the sparse sampling of the shower, the uncertainties in the modeling of the shower development and the large statistical fluctuations in the interactions in the atmosphere, the inferred energy and characteristics of the shower can only determine the relative contribution of groups of elements to the cosmic ray flux. For example,  a heavy nucleus (usually referred to just as ``iron") generates a shower with a relatively higher ratio of muons to electrons that a proton initiated shower of the same energy.

  A good understanding of the composition around the knee would help determining what is the origin of this feature, but this is difficult. The best measurements of the composition of cosmic rays in the knee region come from the Kascade (Karlsruhe Shower Core and Array Detector)    experiment, which observes a steepening of the spectrum first of light elements and successively at higher energies for heavier ones~\cite{knee, Horandel:2006jd} (see Fig.~5). The knee can be explained  by the cutoff  of the galactic proton spectrum due to the leakage of protons from the galaxy.  The galactic magnetic fields are not strong enough to keep protons of this energy bound to the galaxy. Heavier elements are bound to the galaxy up to higher energies, so their cutoff energy increases approximately as their charge $Z$. However, the data  may also be explained if nuclei are accelerated in the supernova shock fronts to maximum energies proportional to $Z$~\cite{Gaisser:2006sf}.  

\begin{figure}[h]
\includegraphics[width=35pc]{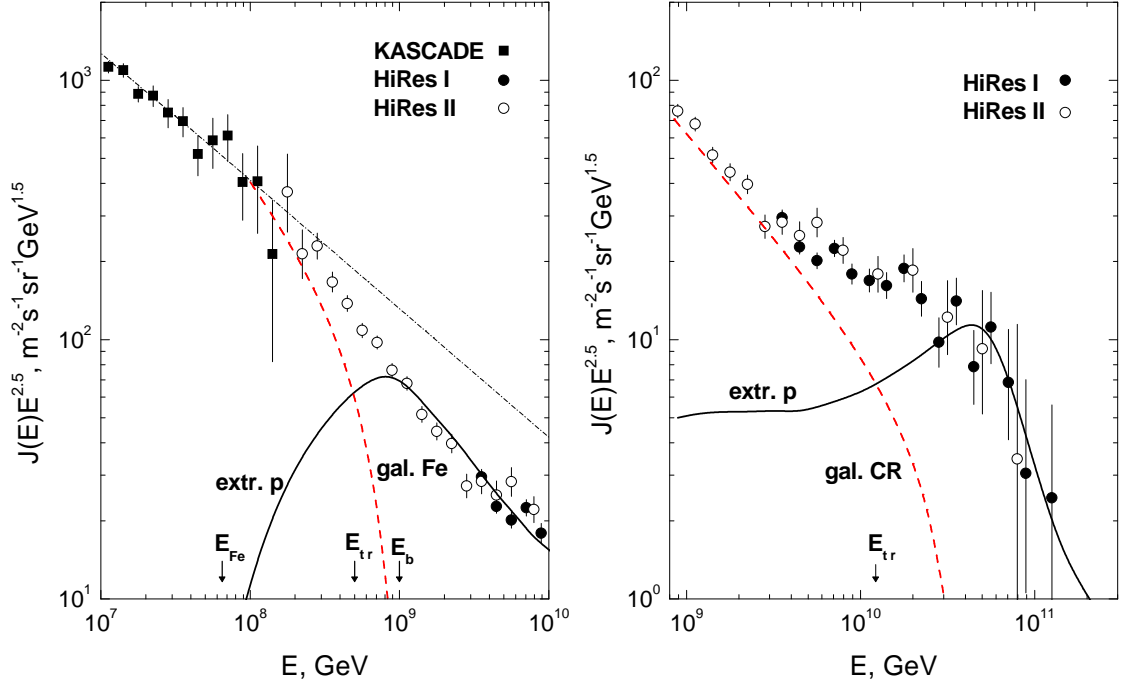}
\caption{\label{fig6} Two possibilities for the crossing of galactic to extragalactic cosmic rays: 3a. (left)
shows the crossing at the ``second knee" $\sim4 \times 10^{8}$ GeV (the ankle is part of a dip due to the interaction of extragalactic protons with the CMB); 3b. (right)  shows the crossing at the ankle,  above 10$^9$ GeV. Figure taken from Ref.~\cite{Aloisio:2007rc}}
\end{figure}
 
  The interpretation of the second knee is linked to that of the ankle. The second knee is sometimes explained by the fall off of the  heaviest elements (including elements heavier than iron) in the galactic cosmic rays~\cite{knee, Horandel:2006jd} (see Fig.~5). The second knee would then be the place in the spectrum where an extragalactic component becomes dominant, as shown in Fig.~6a.
\begin{figure}[h]
\includegraphics[width=35pc]{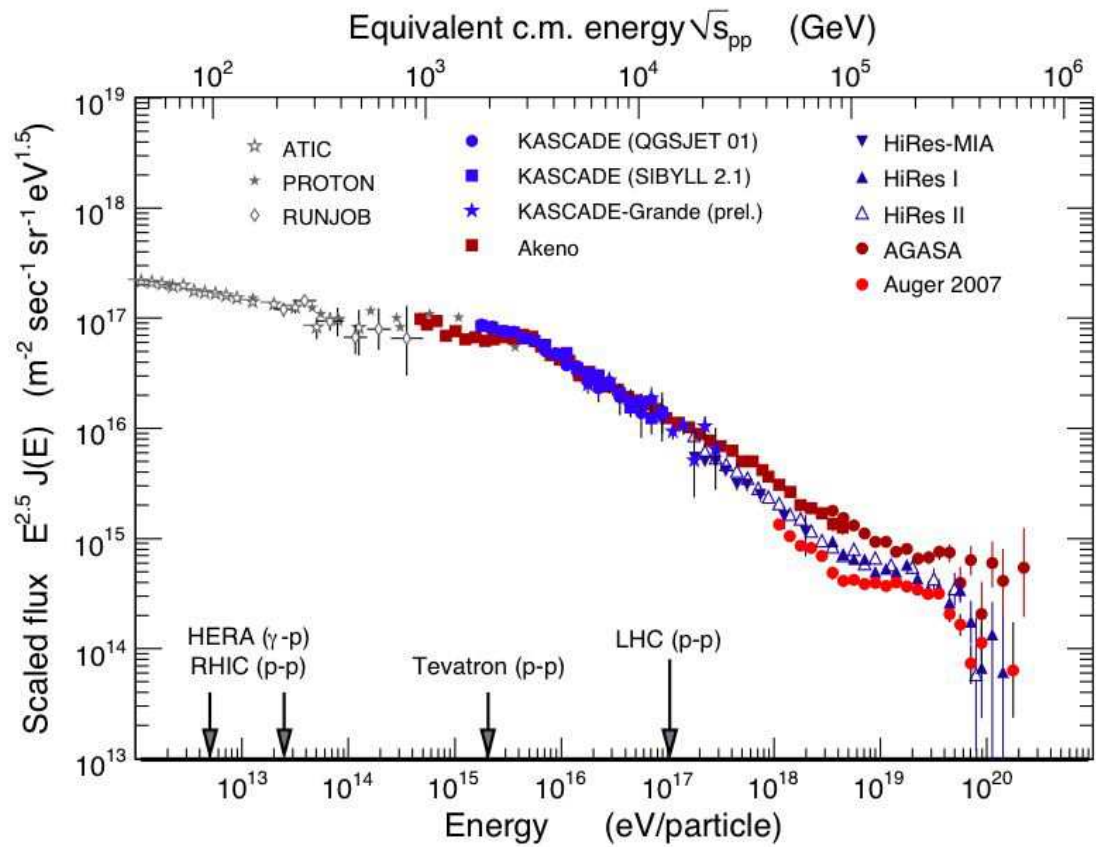}
\caption{\label{fig7} Compilation by R. Engel of the main cosmic rays data up to 2007. The fluxes are multiplied by $E^{2.5}$. The AGASA spectrum does not show a GZK  feature, which is present in the HiRes and 2007 Auger spectra.}
\end{figure}
In this case,  the ankle is an absorption ``dip" in the spectrum between 10$^{18}$ and 10$^{19}$  eV proposed to be caused by electron-positron pair production of extragalactic protons on cosmic microwave background (CMB) photons over large distances~\cite{dip}. This is one of the possible explanations for the ankle. Historically  the ankle was interpreted as the transition from a rapidly falling galactic (iron dominated) component to a flatter spectrum of extragalactic origin subdominant at lower energies. Both interpretations are  shown in Fig.~6b~\cite{Aloisio:2007rc}.  The cosmic rays above 10$^{18}$ eV are called Ultra-High Energy Cosmic Rays (UHECR).

\section{The mystery of the UHECR}
  
Pierre Auger observed   cosmic rays with energies above 10$^{15}$ eV already in the 1930's, and 80 years later their origin is still uncertain. He was first in understanding that extensive air showers are produced by high energy  cosmic rays. Giuseppe Cocconi argued in 1956 that cosmic rays above the ankle
are extragalactic in origin because they cannot be confined by the galactic magnetic field. 
In 1962 John Linsley at Volcano Ranch in new Mexico observed the first 10$^{20}$ eV shower. 

Immediately after the discovery of the microwave background radiation by Penzias and Wilson in 1965, in two separate papers in 1966,  Greisen,  and independently Zatsepin and Kuzmin~\cite{gzk} pointed out that if the UHECR are of extragalactic origin  there should be an absorption feature in their spectrum  at  about 4$\times 10^{19}$ eV, the so called ``GZK cut-off".  In fact, if the UHECR  are extragalactic  their energy is degraded by inelastic collisions   with background photons as they propagate. Protons would loose energy mainly through photopion production in the CMB background  $p\gamma \rightarrow
\Delta^* \rightarrow N\pi$, where the pion carries away $\sim 20\%$ of the
original nucleon energy. The mean free path for this reaction is only
$6$~Mpc. Heavier nuclei mostly loose energy through photodisintegration. For both, protons and heavy nuclei, the energy losses become important above about  4$\times 10^{19}$ eV, thus they could not reach Earth from a distance beyond 50 to 100 Mpc.

\begin{figure}[h]
\includegraphics[width=25pc]{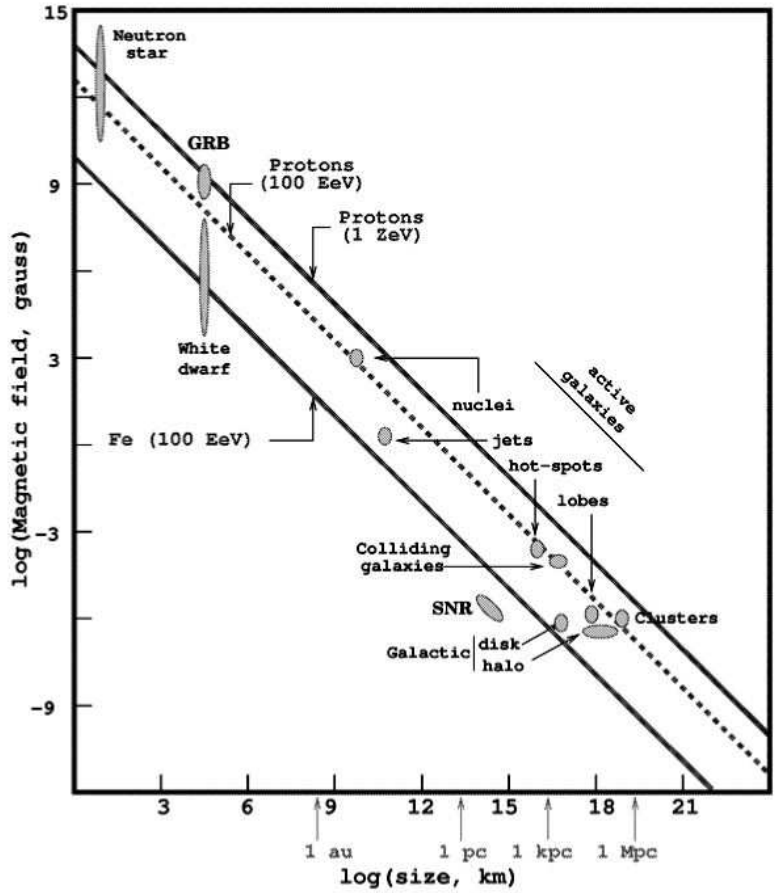}\hspace{2pc}%
\begin{minipage}[b]{10pc} \caption{\label{fig8} ``Hillas plot"  showing the magnetic field versus size of potential UHECR sources. Acceleration of protons to 1$~$ZeV$=10^{21}$ eV, or protons or Fe nuclei to 100 EeV $=10^{20}$ eV require conditions above the respective line.}
\end{minipage}
\end{figure}

 Until recently the existence of the GZK absorption feature was in question. The results of the main UHECR experiments  until a few years ago, AGASA and HiRes, were contradictory (see Fig.7).
The AGASA experiment in Japan, an array of 111 surface detectors (scintillator counters) distributed in 100 km$^2$ with 1 km spacing between them, took data from 1984 to 2003 and  found no GZK cutoff~\cite{agasa}. This gave rise to several models to explain the ``AGASA excess" above 10$^{20}$ eV, in which the UHECR were produced locally  so the cosmological energy absorption would not be important. These so  called ``top-down" models invoked new physics to produce UHECR directly with the high energies required,  as opposed to ``bottom-up" models in which UHECR are accelerated in astrophysical sites. 

  In 1993, the Fly's Eye experiment, consisting  of a fluorescence telescope  in Utah, USA, detected a 3$\times$10$^{20}$ eV event.  The upgrade  of this experiment, the High Resolution Fly's Eye (HIRes), with two fluorescence telescopes took data between 1997 and 2006 and  found a spectrum compatible with the GZK cutoff~\cite{hires}.  
  
  The Pierre Auger Observatory in Argentina~\cite{Auger}, a hybrid experiment which combines  the two detection methods of AGASA and HiRes, was designed  to prove or disprove the existence of the GZK cutoff and to elucidate the mystery of the origin of the highest energy cosmic rays.  It consists of both  surface detectors (water tanks),  SD,  and fluorescence detectors, FD.  The  spectrum of AGASA showing no GZK feature, and those of HiRes and Auger in 2007 compatible with a GZK feature, are shown in Fig.~7.  

Top-down models were introduced not only to produce the highest energy cosmic rays locally and avoid 
the GZK absorption feature, but also  as
an alternative to acceleration models to explain the
 highest energy cosmic rays, which the latter models have
 difficulty explaining.    Only the most extreme sources in the Universe are possible sites for the acceleration to protons or other nuclei at those energies:  Active Galactic Nuclei (AGN), Gamma Ray Bursts (GRB), magnetars (i.e. neutron stars with enormous magnetic fields $B>> 10^{12}$ G), huge lobes and hot spots of radio galaxies, Super Nova Remnants (SNR), colliding galaxies, large shocks in galaxy clusters (for a review see for example Ref.~\cite{astro-sources}, from which
Fig.~8 was taken). The maximum energy at which particles can be accelerated in astrophysical sources is limited  by the acceleration rate and the energy loss rate due to inverse Compton scattering or inelastic scattering. The ``Hillas plot"~\cite{Hillas} (see Fig.~8) shows astrophysical sites for which the gyroradius 
\begin{equation}
r_g \simeq \frac{(E/ 10^{18} {\rm eV}) {\rm Mpc}} { Z (B/ 10^{-9} {\rm G})},
\end{equation}
 of a particle of charge $Z$ and the desired energy $E$, in the estimated magnetic field $B$ of the site, is smaller than the size $L$ of the site. The idea is that only in sites where the particles can be confined for a while, the acceleration could take place. Fig.~8 shows a modern version of the Hillas plot~\cite{astro-sources}, including many more potential production sites than the original version of the plot (from 1984~\cite{Hillas}). As the figure shows, the possible sites where protons could be accelerated diminishes fast as the energy increases above 10$^{20}$ eV.

Top-down models circumvented the need for acceleration of particles by producing them at very high energies. In Super Heavy Dark Matter (SHDM) models, super heavy metastable particles were produced 
in early Universe and remained  at present as part of the dark matter
 of the Universe. These particles (with colorful names such as
`cryptons' or `wimpzillas') could decay or annihilate~\cite{SHDM}  into quarks and leptons etc which in turn produced the
observed UHECR. These would come primordially from within
the dark halo of our galaxy, thus  SHDM models predicted an excess of events towards the galactic center, which has not been found. In these models the maximum energy of cosmic rays was given by the parent particle mass.  In other models, the parent particles instead of being metastable  were instead produced by topological defects, such as cosmic strings or necklaces (for a review see Ref.~\cite{td_review}).  In the ``Z-burst" model~\cite{zburst}, ultra-high energy neutrinos from remote sources  would annihilate at the Z-resonance with relic  background neutrinos near Earth (this model encountered problems with the large flux of ultrahigh energy neutrinos required).  The  Z bosons would then decay,  producing 
secondary protons, neutrinos and photons.
The Z-resonance, which is  the UHECR energy cutoff,
occurs when the energy of the incoming $\nu$ is
$E_{res} = M_Z^2/ 2~m_{\nu}
	= 4\times 10^{21}{\rm eV}({\rm eV}/{m_\nu})$.  The spectra of  the UHECR produced in Top-Down models are determined by the elementary particle physics of Z-boson
 decays  and of QCD fragmentation, which predict photon
 domination of the spectrum at high energies.
\begin{figure}[h]
\includegraphics[width=35pc]{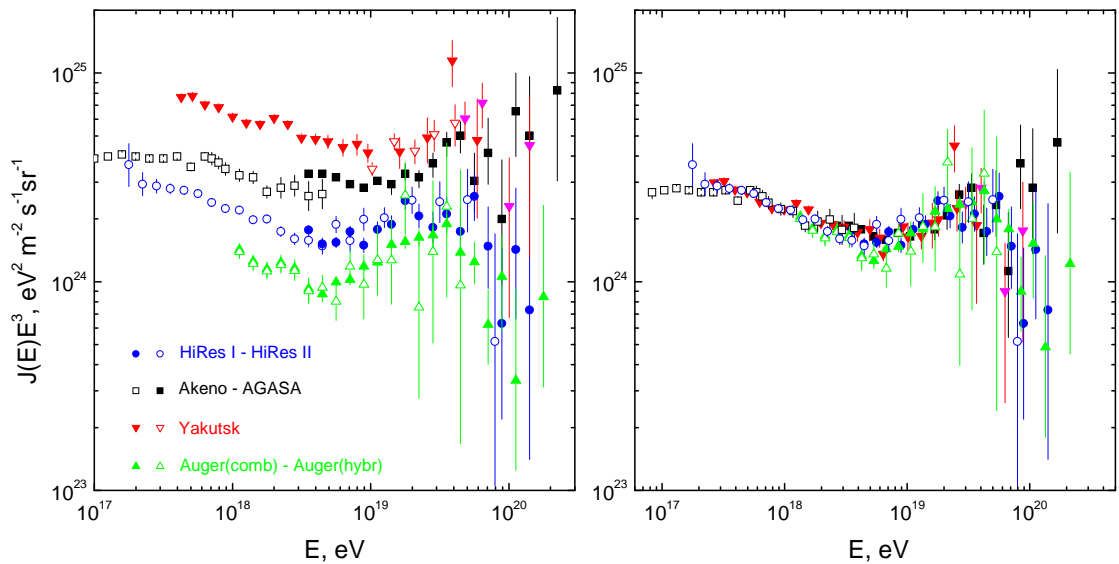}
\caption{\label{fig9} 9a (left panel) Fluxes measured by Yakutsk, AGASA, HiRes and Auger, multiplied by $E^3$. 9b (righ panel) The ``dip" has been used to calibrate the energy of the experiments by factors within their systematic errors. The spectra now coincide, except at the highest energies. See Ref.~\cite{Bere08} for details. }
\end{figure}

The Pierre Auger observatory consists of 1600  water tanks with 1.5 km spacing, distributed  in a 3,000 km$^2$ surface, the surface detector (SD) and four fluorescence detector sites, each with 6  telescopes that overlook the surface array, the fluorescence detector (FD).  The tanks detect the Cherenkov light produced by the electromagnetic and muon component of the air showers.  The telescopes detect the near ultra-violate light from the flourescent emission of Nitrogen induced by the shower.  Auger started operating in 2004 and has been completed a few months ago, in  June 2008. The Auger exposure now is of the order of 10$^4$ km$^2$ sr yr. The FD measures the longitudinal shower distribution and provides a calorimetric measurements of the shower energy, but can operate only in moonless nights, which gives it a 10\% duty cycle. Since the FD telescopes observe the light emitted by the air molecules and not the energy deposited by the shower particles in the air, the energy measurement inferred with the FD depends sensitively on the value adopted for the fluorescence yield, which is  still not well known. 
This is an important source of systematic error. For example, Auger and HiRes use values of the fluorescence yield  that differ by  about 10\%. A better determination of the fluorescence yield is expected soon.  

The SD measures the transverse shower distribution and observes the sky all the time, thus it provides most of the statistics at the  highest energies. About 15\% of the showers observed by Auger are hybrid events, observed both by the SD and the FD. The hybrid events are used to calibrate the energy of the much larger number of SD only events.

 The systematic uncertainly in the energy determination of Auger is at present 22\%.
The fluxes measured by Yakutsk, AGASA, HiRes and Auger are displayed in  Fig. 9a (left panel). The multiplication of the fluxes by $E^3$  exacerbates the importance of the systematic errors in the energy calibration. Fig.9b (right panel) shows how the spectra coincide, except at the highest energies, if the dip feature that appears in all of them is used to calibrate the energy by factors within the systematic errors of the different experiments. The energy calibration factors used to produce Fig. 9b are 1.0, 1.2,  0.75, 0.83 and 0.625 for HiRes, Auger, AGASA, Akeno and Yakutsk respectively (see Ref.~\cite{Bere08} for details).

The Pierre Auger Observatory was optimized  to observe in the energy range between 10$^{18}$ eV and 10$^{20}$ eV, because this is the window in which cosmic rays could reach us from extragalactic sources. The extragalactic magnetic fields are not expected to be larger than 10$^{-9}$ G. Recent realistic
simulations of the expected large scale extragalactic magnetic fields, show that except in the regions
close to the Virgo, Perseus and Coma clusters the obtained magnetic fields are
not larger than 3$\times 10^{-11}$~G~\cite{dolag2004}  (however  see Ref.~\cite{Sigl:2004yk}). We can see then from Eq.~1 that 
with $B < 10^{-9}$ G at energies  $E >10^{18}$ eV protons would only be  weakly  deflected over distances of Mpc.  At energies above 10$^{20}$ eV the GZK cutoff will considerably reduce the flux of UHECR. Thus there is a window between 10$^{18}$ eV and 10$^{20}$ eV to possibly do extragalactic astronomy with charged particles.

The main properties of UHECR Auger is expected to determine are four: 1) the spectrum of the highest energy cosmic rays, in particular to prove or disprove the existence of a GZK cutoff; 2) the composition of those high energy primaries (are they protons, heavy nuclei, photons?); 3) the arrival direction distribution (is there a large scale anisotropy, a correlation with particular sources?);  4) the sources. In the following we address these issues in turn.

\begin{figure}[h]
\begin{minipage}{18pc}
\includegraphics[width=18pc]{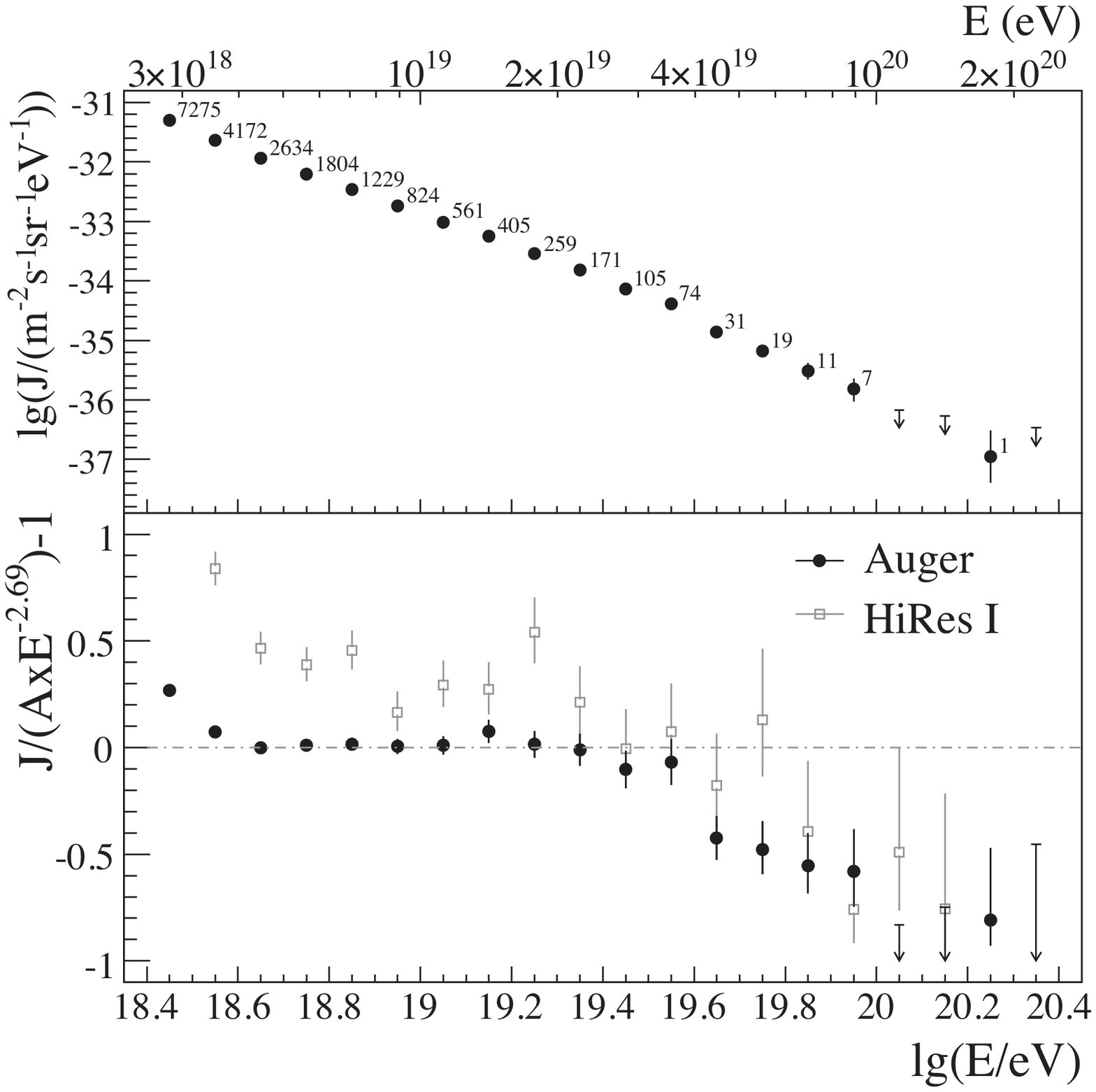}
\caption{\label{fig10} Auger differential energy spectrum~\cite{Abraham:2008ru}. The lower panel shows the fractional differences between  HiRes  and Auger data  and  a spectrum $\sim E^{-2.69}$.}
\end{minipage}\hspace{2pc}%
\begin{minipage}{18pc}
\includegraphics[width=18pc]{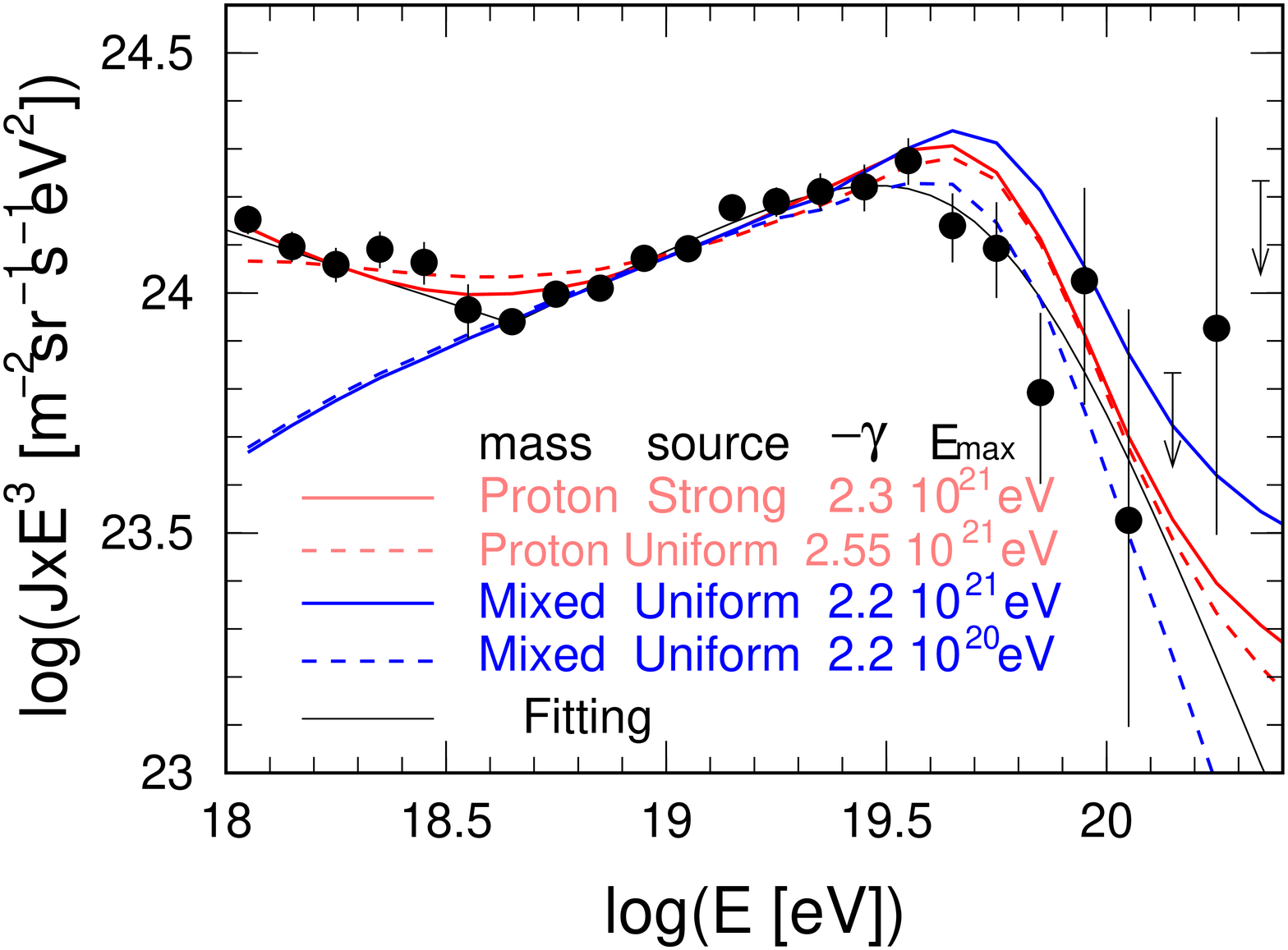}
\caption{\label{fig11} Fits to the auger spectrum with different compositions and injection spectra of primary UHECR~\cite{Allard:2008gj}.}
\end{minipage} 
\end{figure}

The about 2$\times 10^4$ events above 2.5$\times 10^{18}$ eV collected by Auger up to August 2007, with about 4 times the exposure of AGASA and  2 times that of HiRes, have a spectrum compatible with a GZK cutoff at 4$\times 10^{19}$ eV, as shown in Fig.~10~\cite{Abraham:2008ru}. The lower panel of Fig.~10 shows the fractional differences between the HiRes and Auger data  at the highest energies and  a power law  $\sim E^{-2.69}$ which fits the spectrum below 4$\times 10^{19}$ eV.  The number of events expected if this power law spectrum were to continue to energies above  4$\times 10^{19}$ eV or above
1$\times 10^{20}$ eV would be 167$\pm$3 and 35$\pm$1 respectively, while Auger observed 69 and 1 respectively.  This amounts to a 6$\sigma$ significance of the flux suppression at the GZK energy, which is compatible with what was found by HiRes.

The spectrum cannot be used to determine the composition of UHECR, since models with pure proton primaries and mixed proton-iron primaries all produce good fits to the spectrum. Fig.11 shows several fits to the spectrum of Auger with proton primaries and mixed composition primaries and several primary spectral indices $\gamma$ and maximum energies $E_{\rm max}$ characterizing the spectrum emitted by the sources~\cite{Allard:2008gj}.  Even the extreme unrealistic assumption of having  pure iron emitted at the sources provides  good fits to the Auger spectrum~\cite{Arisaka:2007iz}. It is also not possible either to determine from the spectrum if the observed suppression is due to the GZK effect or to the energy cut of the spectra of the cosmic rays emitted at the sources. 
\vspace{1.0cm}

\hspace{1cm} \begin{figure}[h]
\includegraphics[width=20pc]{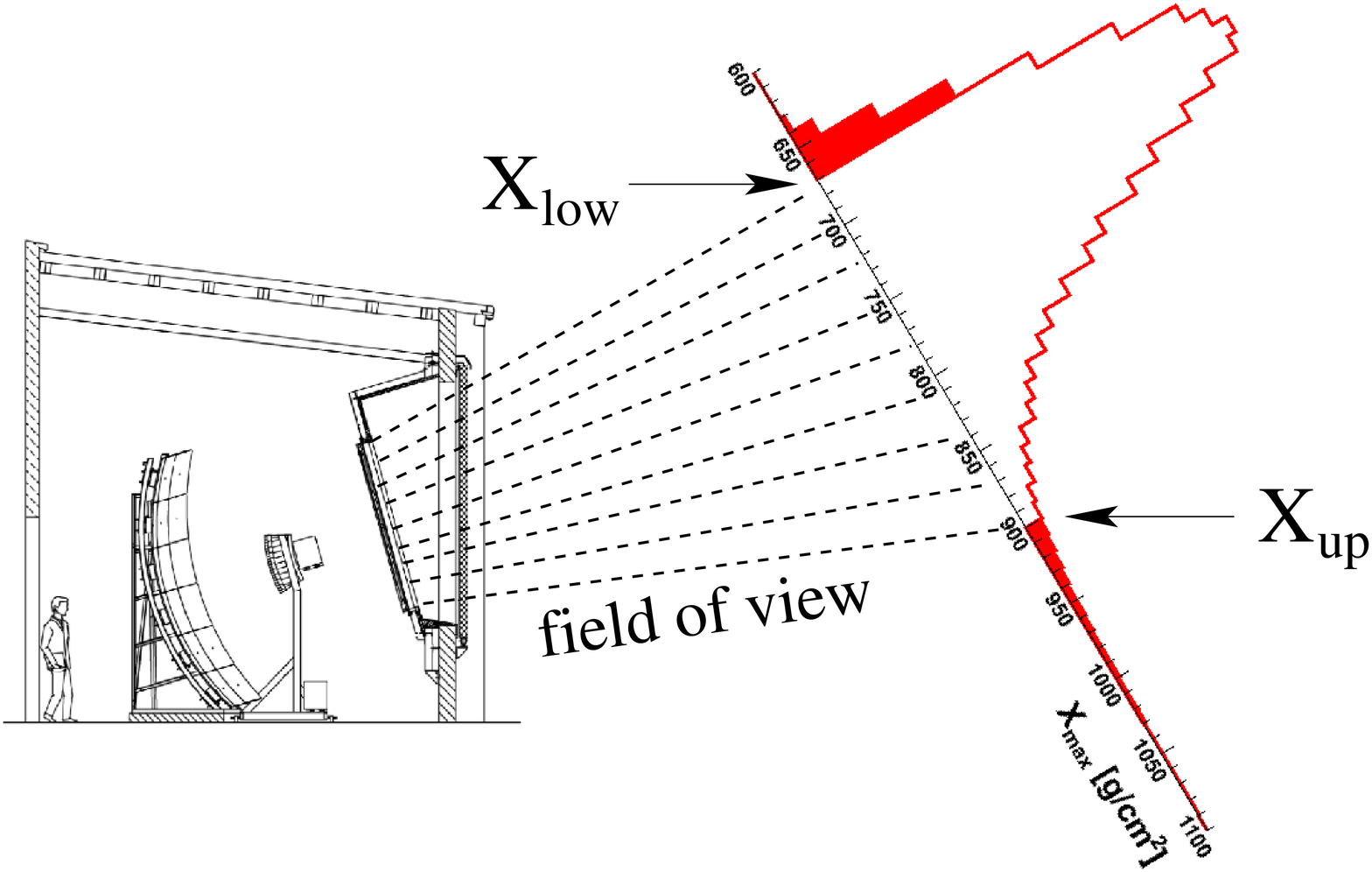}\hspace{2pc}%
\begin{minipage}[b]{15pc} \caption{\label{fig12} Illustration of the observation of the longitudinal shower development with the Auger  FD. The maximum defines $X_{\rm max}$. From Ref.~\cite{Unger:2007mc}}
\end{minipage}
\end{figure}

The main tool to study the mass composition of cosmic rays is the measurement of the longitudinal profile of the showers. Fig.~12 shows an illustration of the measurement of the longitudinal shower development
observed with the fluorescence detectors of Auger. For a given energy,  heavy nuclei interact earlier in the atmosphere than protons (nuclei interact as a collection of nucleons). Photon initiated showers are much more penetrating in the atmosphere than hadrons, up to energies higher than 10$^{20}$ eV at which photons produce a ``preshower"   (through interaction with the magnetic field of the Earth) before entering into the atmosphere. Thus the column depth of air at which the shower profile is maximum, $X_{\rm max}$, as measured with the FD, is well correlated with the primary particle mass. This can be seen in Fig.~13 where the expected $X_{\rm max}$ as function of energy for photon, proton and iron initiated showers are shown, as predicted by several simulation programs. The value of the $X_{\rm max}$ for protons is about 100 g cm$^{-2}$ larger than for iron. The FD detector of Auger can measure $X_{\rm max}$ with a systematic uncertainty of about 15 g/cm$^2$. Fig.~13 also shows a compilation of the earlier data on $X_{\rm max}$ as function of the energy, for energies above 10$^{14}$ eV. It shows a composition compatible with proton dominance at the highest energies. Except for one data point of HiRes, the measurements reached 2$\times 10^{19}$ eV.

\begin{figure}[h]
\includegraphics[width=30pc]{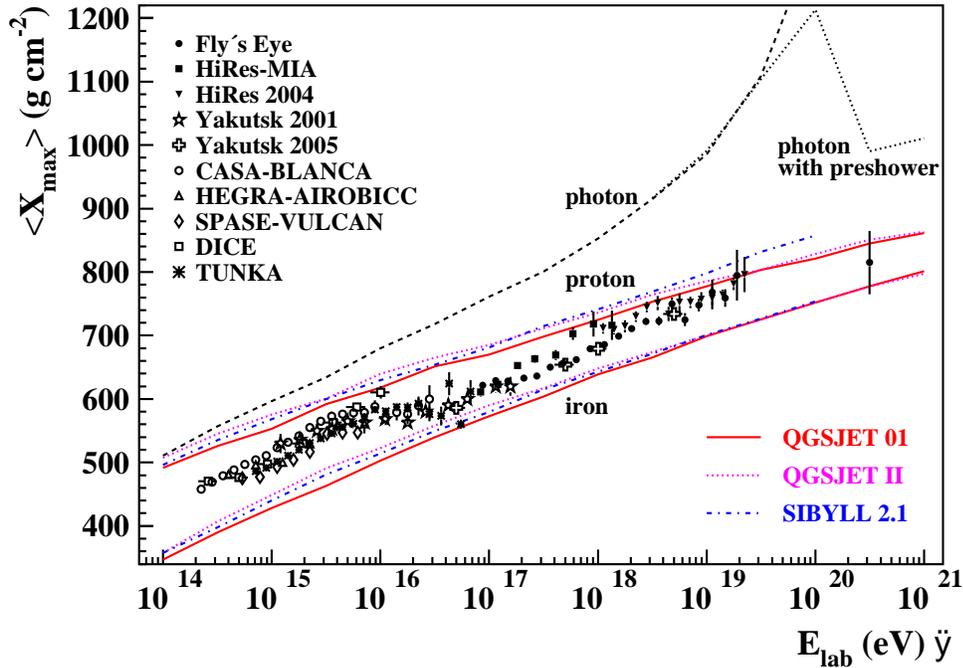}
\caption{\label{fig13} Compilation of earlier measurements of the mean depth of shower maximum $X_{\rm max}$ as function of energy, compared with the predictions of various simulations programs for photon, proton and iron initiated showers (see for example Ref.~\cite{Matthiae}). }
\end{figure}

Auger data on  $X_{\rm max}$~\cite{Unger:2007mc} are presented in Fig.~14 together with the predictions for proton and iron initiated showers from several shower simulation programs. Although with  low statistics the Auger data show a change of regime (a change in slope, called ``elongation rate")  at 2$\times 10^{19}$ eV towards a heavier composition. At the highest energies the composition  is intermediate between protons and iron, compatible with a mean mass number of about 5. The errors shown in the figure are statistical. The systematic uncertainty is  $\pm$ 15 g/cm$^2$. The HiRes collaboration did not find the change in slope at 2$\times 10^{19}$ eV present in the Auger data, and its composition measurements seem compatible with proton dominance to the highest energies  (see Fig.16). 
\begin{figure}[h]
\includegraphics[width=30pc]{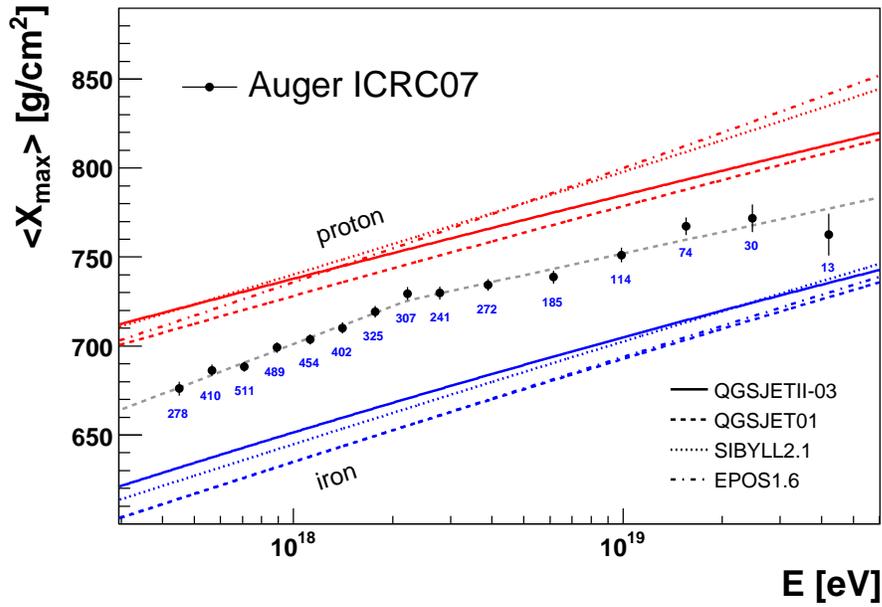}
\caption{\label{fig14} Mean depth of shower maximum $X_{\rm max}$ as function of energy measured by Auger~\cite{Unger:2007mc} compared with what is expected from proton and iron initiated showers. The last data point seems to indicate a heavier composition above 2$\times 10^{19}$ eV.}
\end{figure}
\begin{figure}[h]
\hspace{0.5cm}\includegraphics[width=28pc]{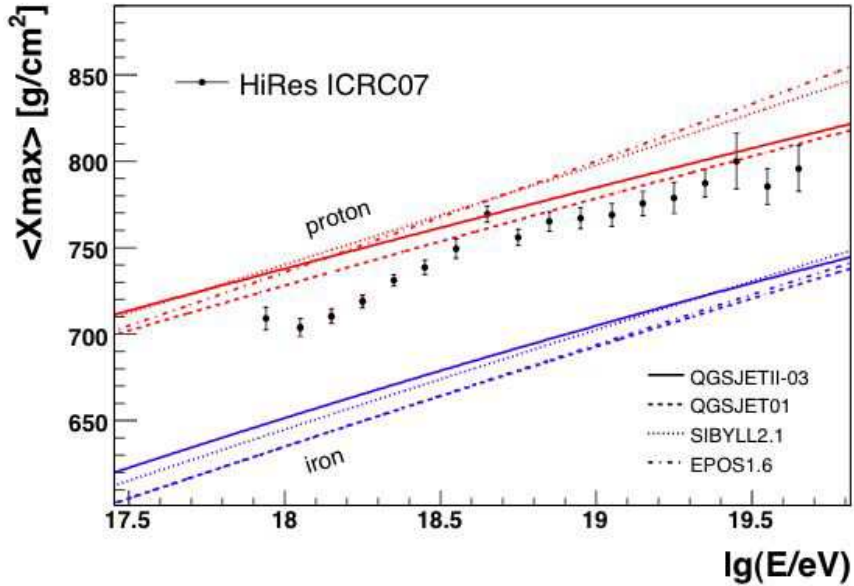}
\caption{\label{fig15} Mean depth of shower maximum $X_{\rm max}$ as function of energy measured by HiRes~\cite{HiRes-elongation} (fig. from Ref.~\cite{Engel}) compared with what is expected from proton and iron initiated showers. No change in the slope is observed at 2$\times 10^{19}$ eV. }
\end{figure}

Auger has obtained stringent upper bounds on photons as UHECR~\cite{Abraham:2006ar, Aglietta:2007yx}.  UHECR photons with energy above 10$^{19}$eV are one of the key observables to distinguish Top-Down from Bottom-Up models (see for example Ref.~\cite{Gelmini:2005wu}).
Because in the process of hadronization pions are produced more copiously than protons, and neutral pions decay into photons, all Top-Down models produce more photons than protons (and no heavy nuclei). Also ``GZK" photons, those produced in the decay of photo-produced $\pi^0$ in the GZK process must always be present at some level, although the level may be extremely low. The photon fraction could be
as small at 10$^{-4}$~\cite{Gelmini:2007sf, Gelmini:2007jy}. Finding GZK photons among the UHECR  would help understanding the primary hadronic spectrum which produced them  and the intervening extragalactic backgrounds with which they interact, the radio background and magnetic fields. The flux of GZK photons is related to that of ``cosmogenic" neutrinos, those neutrinos produced in the decay of the photo-produced charged pions in the GZK process.   Neutrinos and photons  produced as secondaries of the GZK process would give complementary information since neutrinos come without absorption from the whole production volume, while photons can reach us only from less than 100 Mpc.  Photons with  energy $> 10^{19}$ eV pair-produce electrons and positrons
on the radio background and cannot reach Earth from beyond 10 to 40
Mpc (although the photon energy-attenuation length is not well known,
due to the uncertainties in the spectrum of the absorbing radio
background). 
\begin{figure}[h]
\begin{minipage}{19pc}
\includegraphics[width=19pc]{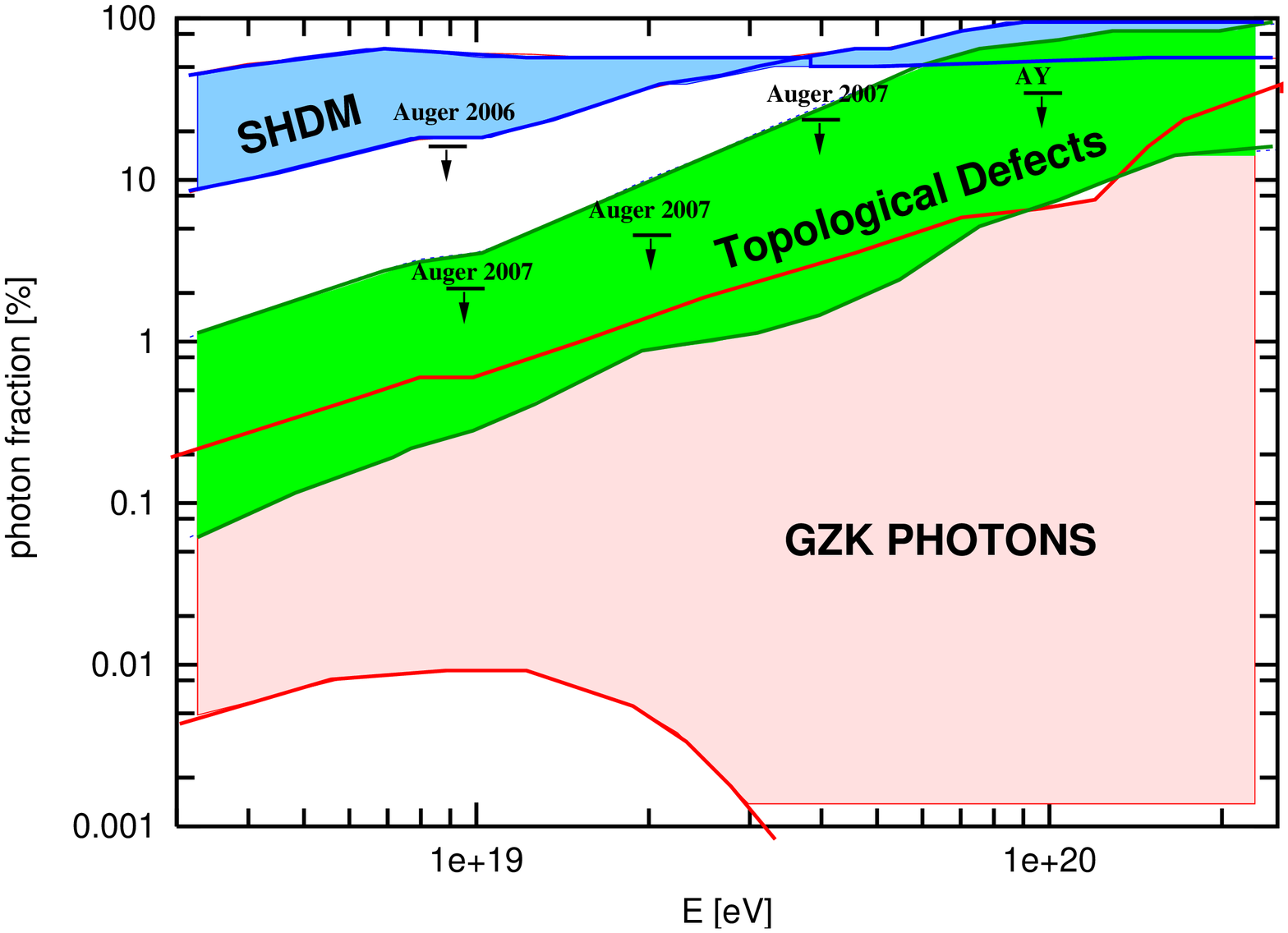}
\caption{\label{fig16} Range of photon fractions of UHECR given as \% of the integrated total flux for several Top-Down models and GZK photons. The 2006 and 2007 Auger  and AGASA-Yakutsk (AY) upper bounds are also shown. From Ref.~\cite{Gelmini:2005wu}.}
\end{minipage}\hspace{2pc}%
\begin{minipage}{18pc}
\vspace{-0.7cm}
\includegraphics[width=18pc]{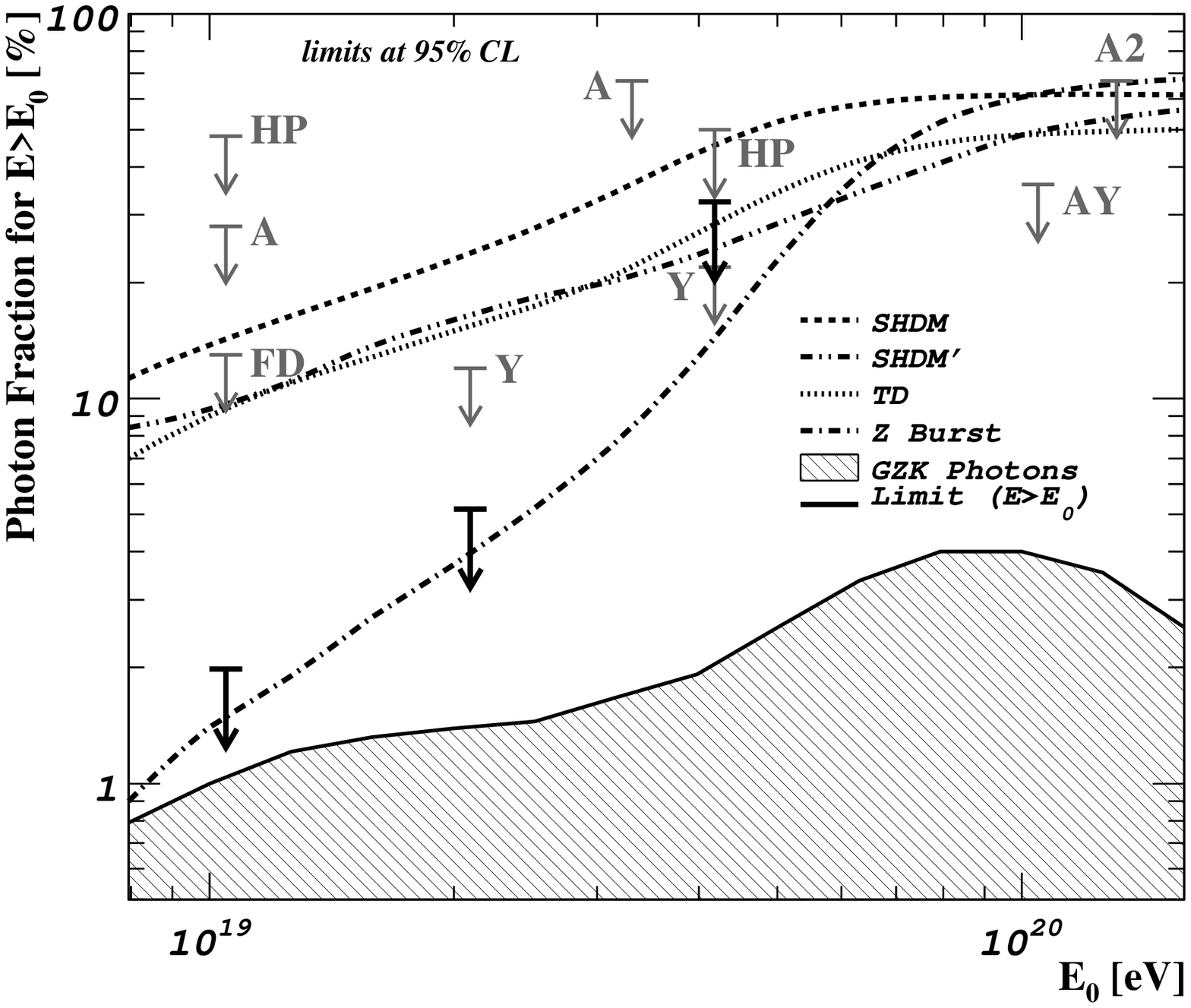}
\caption{\label{fig17} Same as Fig.~13, but with several characteristic Top-Down models and other experimental upper bounds. See Ref.~\cite{Aglietta:2007yx} for details.}
\end{minipage} 
\end{figure}

At energies below 10$^{20}$ eV photons produce in the atmosphere deeper showers  than protons (due to the Landau-Pomeranchuk-Migdal (LPM) effect). At higher energies the photons interact with the magnetic field of the Earth before entering the atmosphere and produce several electrons below the LPM threshold (what is called ``preshowering") but the statistics of Auger  is too low at those energies.  Bounds based on $X_{\rm max}$ measurements using FD data were initially used to set an upper bound on photons~\cite{Abraham:2006ar}.  Better bounds were obtained later~\cite{Aglietta:2007yx}
by using two parameters, the shower-front radius and the rise time of the signal, that can be measured with the SD. Late developing showers  have smaller shower-front radius (the shower-front flattens out as the shower develops).    Photon initiated showers have a lower muon content, more low energy particles, and thus longer rise-time of the signal. With these two parameters, the high statistics SD could be used to search for photons. Auger found no events compatible with what is expected from a photon primary. As a result it has obtained stringent upper bounds shown in Figs. 16 and 17. The range of photon fractions in the integrated flux for some Top-Down models and for GZK photons from Ref.~\cite{Gelmini:2005wu} together with the 2006 and 2007 Auger  and AGASA-YAKUTZK (AY)~\cite{AgasaYakutskLimit}  upper bounds are shown in Fig.~16.  The Auger bounds together with other upper bounds on the photon fraction and characteristic Top-Down models are shown in Fig.~17~\cite{Aglietta:2007yx}. These photon bounds provide an argument independent of possible correlation with astrophysical sources against most Top-Down models.

If the suppression of the cosmic rays spectrum at 4$\times 10^{19}$ eV is due to the GZK effect, namely to the interactions with the CMB photons, then the cosmic rays with higher energies should come from  nearby sources. For example, above 6$\times 10^{19}$ eV 90\% of the protons should come from less than about 200 Mpc away, while 50\% should come from less than 90 Mpc away. Thus, the arrival directions of the highest energy cosmic rays should correlate with the distribution of visible matter nearby, which is very inhomogeneus.  The angular resolution of Auger above 1$\times 10^{19}$ eV is less than one degree, while the deflection of protons or nuclei in the $\mu$G magnetic field of our galaxy, at those energies is  expected to be about 10$^o Z (10^{19} eV/E)$. Auger has, in fact, found a correlation with extragalactic astrophysical sources~\cite{Cronin:2007zz, Abraham:2007si}.

Auger reported  a correlation between the arrival directions
 of UHECR with $E> 5.7 \times 10^{19}$ eV
 and  the positions of nearby active galactic nuclei (AGN) with redshift $z \leq 0.017$ (i.e. distance $<$71 Mpc): 20 out of 27 cosmic ray events were found to correlate  with at least one of 442  AGN selected form a particular catalog, the Veron-Veron-Cetty catalog (292 within the field of view of Auger) located within a circle of 3.2$^o$ around the event arrival direction (only 5.6 are expected to do so if the flux were isotropic). The Auger collaboration found a statistically significant correlation with a smaller amount of data in May 2006 with very similar energy threshold and angular distance. Fixing these parameter a-priori it studied the subsequent data until August 2007 when the original correlation was confirmed to a pre-determined confidence level in the additional data set alone.

\begin{figure}[h]
\includegraphics[width=27pc]{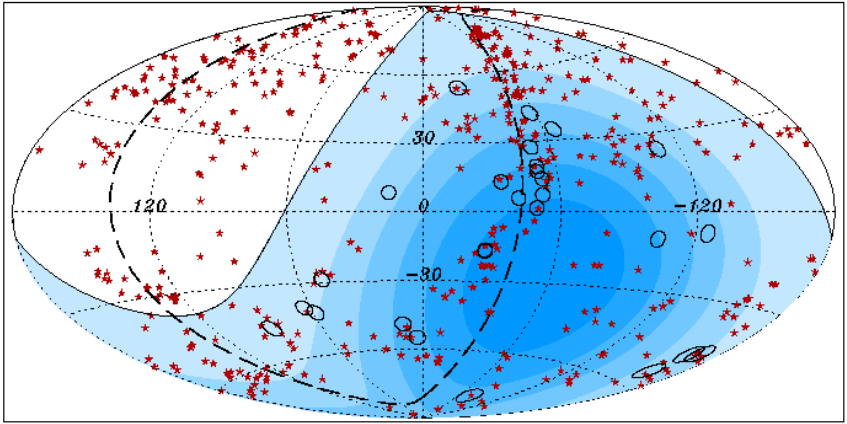}\hspace{2pc}%
\begin{minipage}[b]{10pc} \caption{\label{fig18} Map in galactic coordinates of the positions of the AGN
within 71 Mpc used in the correlation search of Auger (red stars) and circles of 3.2$^o$ centered at the arrival directions of the 27 events with $E> 5.7 \times 10^{19}$ eV observed by Auger until August 2007. The shading indicate levels of exposure~\cite{Cronin:2007zz, Abraham:2007si}.}
\end{minipage}
\end{figure}

 The map of the arrival directions and of the AGN positions is shown in Fig.~18. There is 
a clear alignment of several of the correlating UHECR events close to the supergalactic plane, indicated in the figure with  a dashed line.  Two of the events have arrival directions less than 3$^o$ away from the nucleus of Centaurus A, one of the closest AGN, at 3.4 Mpc away. Cen A displays jets and radio lobes which extend over a scale of about 10$^o$ along the supergalactic plane, and a variable compact radio  nucleus. The two events mentioned correlate with the nucleus position while several lie in the vecinity of the radio lobe extension along the supergalactic plane.  In general, AGN with prominent radio lobes are rare and do not follow the observed spatial distribution. So far there are only three other doublet events (i.e. the same AGN associated with two UHECR), one of which is close to and other two which are outside the supergalactic plane.  

 The most prominent radio galaxy within 70 Mpc, M87 in the Virgo cluster, does not correlate with any observed  event above 5.7 $\times 10^{19}$ eV.  There is a  lack of events in the region of the Virgo cluster which seems to be present also in the data of the HiRes Collaboration.  May be large magnetic fields in the direction of Virgo could explain the lack of events in that direction.
 
 The HiRes collaboration did not confirm a correlation of UHECR with AGNs~\cite{Abbasi:2008md}): out of 13 events with $E> 5.7 10^{19}$  only two correlated with an AGN within 3.2$^o$  (and 3.2  events were expected just by chance).  HiRes observed mostly the Northern sky and Auger the Southern, could they be significantly different?

 Given the small statistics of the sample and the negative results of HiRes, the  correlation  found by Auger does not prove that the sources of UHECR are AGNs. The sources may be any astrophysical objects with a space distribution similar to AGNs. Are AGN the sources or just tracers of the sources?
 
  Nuclei are attenuated by photodisintegration in the intergalactic medium and intermediate mass nuclei (A $\simeq 20 -40$) would have to reach us from distances shorter than 70 Mpc. Heavier nuclei, close to Fe  have horizons comparable to that of protons but they would be deflected in the magnetic field of our galaxy by more than 3$^o$.  Thus the correlation indicates that the primary UHECR are most likely protons.  However, in the case the primaries are protons,
  the actual sources of  the UHECR may be further out than 70 Mpc. We mentioned above that  the ``GZK horizon" (defined as the distance from  Earth which contains the sources  that produce 90\% of the protons which arrive with energies above a given threshold)  should be about 200 Mpc for protons of energy above 6 $\times 10^{19}$ GeV, while it would be 70 Mpc for about  9 $\times 10^{19}$ GeV.  This may indicate that the energy calibration of Auger should be shifted upwards by 30\% or  that there are accidental correlations with AGNs that are closer than the actual UHECR sources.  
  
  The correlation with AGNs  found by Auger seems to indicate that the highest energy cosmic rays are
  protons, while the composition data of Auger suggest a change towards a heavier composition above $2\times 10^{19}$ eV. Which of these two pieces of information is correct? There may also be a problem with the programs used to simulates the showers and determine if they are proton or iron like. Data from LHCf  are needed to improve them.
  
  Clearly more data are needed to answer all these questions, and they are coming. There is an infield with extra muon counters  in construction in the Pierre Auger Observatory in Argentina. Also a new experiment, the Telescope Array (with a low energy extension called TALE) is being built in Utah (USA). This Auger Observatory in Argentina is the South part of the two-part  project initially proposed for Auger. There is a current proposal for an Auger-North Observatory in Colorado (USA) with 4$\time 10^3$ surface detectors occupying an area of 2$\times 10^4$ km$^2$. There is also  the JEM-EUSO mission in preparation, the Extreme Universe Space Observatory that will be hosted in the Japanese Experimental Module  of the International Space Station, that will study very energetic cosmic ray showers from space.
  
\section{Concluding remarks}  
 Many advances have occurred in recent years in our understanding of the origin and propagation of cosmic rays but many fundamental open question remain. In the GeV to TeV energy range,  the solar modulation of cosmic rays and the solar wind need to be better understood, and we need to understand if the PAMELA/ ATIC excess is in fact there and it is not an instrumental issue, and if so if it is due to a pulsar nearby or the dark matter annihilation. The Fermi Space telescope data expected soon will greatly help in this respect.  In the 10$^4$-10$^8$  GeV energy range, the cosmic ray composition, the galactic sources, the propagation of cosmic rays within the galaxy and the transition from the galactic to  extra-galactic cosmic rays (at the second knee or at the ankle?) need to be better understood. At energies above 10$^9$ GeV charged particle extragalactic astronomy has started!  but we need yet to identify the sources, understand the composition of the cosmic rays, the mechanisms of acceleration  at the sources and the intervening backgrounds and magnetic fields.




\section*{References}

\begin{thebibliography}{9}

\bibitem{CR} Cronin J, Gaisser T K and Swordy S P 1997 {\it Sci. Amer.} {\bf 276} 44.


\bibitem{Adriani:2008zr}
  Boezio M {\it et al.} 2008
  arXiv:0810.3508 [astro-ph];
  O.~Adriani {\it et al.}  [PAMELA Collaboration] 2008
  arXiv:0810.4995 [astro-ph].

  
 \bibitem{HEAT}
  Barwick  S W {\it et al.}  [HEAT Collaboration] 1997
  {\it Astrophys.\ J.}   {\bf 482}  L191
  [arXiv:astro-ph/9703192];
 Beatty J J {\it et al.} 2004
  {\it Phys.\ Rev.\ Lett.} {\bf 93} 241102
  [arXiv:astro-ph/0412230].
  
  
\bibitem{AMS}
  Aguilar M {\it et al.}  [AMS-01 Collaboration] 2007
  {\it Phys.\ Lett.}  B {\bf 646} 145
  [arXiv:astro-ph/0703154].  
  
    
  \bibitem{ATIC}
  Chang J {\it et al.} 2008
  {\it Nature} {\bf 456} 362.

  \bibitem{HESS} Aharonian F A {\it et al.} 2008  {\it Phys. Rev. Lett.} {\bf 101} 261104
  
  

  
  
  \bibitem{pulsars} Aharonian F A, Atoyan A and Volk H J 1995 {\it Astron. Astrophys.} {\bf 294} L41;
  Hooper D, Blasi P and Serpico PD 2008
 {\it JCAP} {\bf 0901} 025
  [arXiv:0810.1527 [astro-ph]];
 Yuksel H, Kistler M D and Stanev T 2008
  arXiv:0810.2784 [astro-ph];
  Profumo S 2008
  arXiv:0812.4457 [astro-ph].
  
  \bibitem{Hall-Hooper} Hall J and Hooper D 2008  [arXiv:0811.3362 [astro-ph]].
  
\bibitem{Bergstrom:2008gr}
 Bergstrom L,  Bringmann T and Edsjo J,  2008
  {\it Phys.\ Rev.\  D} {\bf 78}, 103520 
  [arXiv:0808.3725 [astro-ph]]. 

  \bibitem{Vogelsberger} Vogelsberger M, Helmi A, Springel V, White S, Wang J, Frenk C, Jenkins A, Ludlow A and Navarro J  2008  arXiv:0812.0362 [astro-ph].


  
\bibitem{Hisano:2004ds}
  Hisano J, Matsumoto S, Nojiri  M M and Saito O 2005
 {\it Phys.\ Rev.\  D} {\bf 71} 063528
  [arXiv:hep-ph/0412403].


\bibitem{Adriani:2008zq}
  Adriani O {\it et al.} 2008
  {\it Phys.\ Rev.\ Lett.}  {\bf 102} 051101
  [arXiv:0810.4994 [astro-ph]].
  
\bibitem{Kuhlen:2008aw}
  Kuhlen M, Diemand J and Madau P, 2008
  arXiv:0805.4416 [astro-ph].
  
  
\bibitem{Horandel:2006jd}
  Horandel J R 2007
 {\it Mod.\ Phys.\ Lett.} A {\bf 22} 1533
  [arXiv:astro-ph/0611387]

  
\bibitem{Gaisser:2006sf}
  T.~K.~Gaisser and T.~Stanev,
  Nucl.\ Phys.\  A {\bf 777} (2006) 98.
  
  \bibitem{knee} 
  Horangel J R 2003 {\it Astropart. Phys} {\bf 19} 193-220;
  Ulrich H {\it et al.} 2007
  {\it AIP Conf.\ Proc.}  {\bf 928} 31.
  
  
  
  \bibitem{dip} Berezinsky V, Gazizov AZ, Grigorieva S I  2004 {\it Nucl. Phys. Proc.
  Suppl.} {\bf 136} 147 [astro-ph/0410650]; Berezinsky V, Gazizov AZ, Grigorieva S I  2005 {\it Phys. Lett.} B {\bf 612} 147 [astro-ph/0502550]; Berezinsky V, Gazizov AZ, Grigorieva S I  2006 {\it Phys. Rev.} D  {\bf 74} 043005
  
  
\bibitem{Aloisio:2007rc}
  Aloisio R, Berezinsky V, Blasi P and Ostapchenko S 2008
  {\it Phys.\ Rev.}  D {\bf 77} 025007
  [arXiv:0706.2834 [astro-ph]].
  
 \bibitem{gzk}
Greisen K 1966
{\it Phys.\ Rev.\ Lett.} {\bf 16}, 748.
Zatsepin G T and Kuzmin V A 1966
{\it JETP Lett.} {\bf 4}, 78 
[{\it Pisma Zh.\ Eksp.\ Teor.\ Fiz.}  {\bf 4} 114].


\bibitem{agasa}
Takeda  M{\it et al.} 1998
{\it Phys.\ Rev.\ Lett.} {\bf 81} 1163
[astro-ph/9807193];
Hayashida  N {\it et al.} 2008
astro-ph/0008102;
{\sf http~://www-akeno.icrr.u-tokyo.ac.jp/AGASA/}.

\bibitem{hires}
Abbasi R U {\it et al.}  [High Resolution Fly's Eye Collaboration] 2004
{\it Phys.\ Rev.\ Lett.}  {\bf 92}, 151101
[arXiv:astro-ph/0208243];
{\sf http~://hires.physics.utah.edu/}.
 
  
  
\bibitem{Auger} Pierre Auger Observatory, http://www.auger.org.  
 
   \bibitem{astro-sources} Ostrowsky M 2002 {\it Astrprt. Phys.} {\bf 18} 229-236
  
  \bibitem{Hillas} Hillas A M 1984 {\it A R A and A } {\bf 22} 425
  
  
  \bibitem{SHDM}
Berezinsky V , Kachelriess M and Vilenkin A 1997
{\it Phys.\ Rev.\ Lett.}\  {\bf 79} 4302;
Kuzmin V A  and Rubakov V A 1998
{\it Phys.\ Atom.\ Nucl.}\  {\bf 61} 1028
[{\it Yad.\ Fiz.}\  {\bf 61} 1122];
Birkel M and Sarkar S 1998
{\it Astropart.\ Phys.}\  {\bf 9}  297; 
Blasi P, Dick R and Kolb E W 2002
{\it Astropart.\ Phys.}\  {\bf 18} 57;
Aloisio R, Berezinsky V and Kachelriess M 2004
  {\it Phys.\ Rev.}\ D {\bf 69} 094023.


\bibitem{td_review}
  Bhattacharjee P and Sigl G 2000 {\it Phys.\ Rept.}\  {\bf 327} 109.
  
  \bibitem{zburst}
Weiler T J 1982 {\it Phys.\ Rev.\ Lett.}\   {\bf 49} 234  and 1984
{\it Astrophys.\ J.}\  {\bf 285}  495;
Fargion D, Mele B and Salis A 1999
{\it Astrophys.\ J.}\   {\bf 517} 725;
Weiler T J  1999 {\it Astropart.\ Phys.}\   {\bf 11} 303.

\bibitem{Bere08}
Berezinsky V 2008
 {\it J.\ Phys.\ Conf.\ Ser.}\  {\bf 120}, 012001
  [arXiv:0801.3028 [astro-ph]].

\bibitem{dolag2004}
Dolag K, Grasso D, Springel V and Tkachev I 2004
{\it JETP Lett.}  {\bf 79}, 583 
[{\it Pisma Zh.\ Eksp.\ Teor.\ Fiz.}  {\bf 79}, 719 ]
[arXiv:astro-ph/0310902]; 
Dolag K, Grasso D, Springel V and Tkachev I 2005
{\it JCAP} {\bf 0501} 009 
[arXiv:astro-ph/0410419].


\bibitem{Sigl:2004yk}
Sigl G, Miniati F and Ensslin TA 2003
{\it Phys.\ Rev.\ D} {\bf 68}, 043002
[arXiv:astro-ph/0302388];
%
%
Sigl G, Miniati F and Ensslin TA 2004
{\it Phys.\ Rev.\ D} {\bf 70}, 043007 
[arXiv:astro-ph/0401084].


\bibitem{Abraham:2008ru}
  Abraham J {\it et al.}  [Pierre Auger Collaboration] 2008
 {\it  Phys.\ Rev.\ Lett.}  {\bf 101} 061101
  [arXiv:0806.4302 [astro-ph]].





\bibitem{Allard:2008gj}
 Allard D, Busca N G, Decerprit G, Olinto A V and Parizot E 2008
  {\it JCAP} {\bf 0810} 033 
  [arXiv:0805.4779 [astro-ph]].



\bibitem{Arisaka:2007iz}
  Arisaka K, Gelmini G B, Healy M, Kalashev  O and Lee J 2007
 {\it JCAP} {\bf 0712} 002
  [arXiv:0709.3390 [astro-ph]].



\bibitem{Unger:2007mc}
  Unger M [The Pierre Auger Collaboration] 2007
  arXiv:0706.1495 [astro-ph]
  
  
  \bibitem{Matthiae}
  Matthiae G 2008
  arXiv:0807.1024 [astro-ph].
  
  \bibitem{HiRes-elongation} Fedorova Y [HiRes collaboration] 2007
  {\it Proceed. of the 30th International Cosmic Ray Conference} {\bf 4} 463-464;
  Sokolsky P  [HiRes Collaboration] 2008
  {\it Nucl.\ Phys.\ Proc.\ Suppl.}\  {\bf 175-176} 207  
  
 \bibitem{Engel} Engel R [The Pierre Auger Collaboration] 2008
 talk at the {\it ISVHECRI Workshop}, Paris,  Sep. 1-5.
  
  
\bibitem{Abraham:2006ar}
  Abraham  J {\it et al.}  [Pierre Auger Collaboration] 2007
  {\bf Astropart.\ Phys.}  {\bf 27} 155 
  [arXiv:astro-ph/0606619].
  
  
 
\bibitem{Aglietta:2007yx}
  Abraham  J {\it et al.}  [Pierre Auger Collaboration] 2008
  {\it Astropart.\ Phys.}  {\bf 29}, 243 
  [arXiv:0712.1147 [astro-ph]].
 
  
  
  
  
\bibitem{Gelmini:2005wu}
Gelmini G , Kalashev O and Semikoz D V 2008
 {\it J.\ Exp.\ Theor.\ Phys.} {\bf 106}, 1061 
  [arXiv:astro-ph/0506128].
  
   
\bibitem{Gelmini:2007sf}
Gelmini G , Kalashev O and Semikoz D V 2007
  {\it Astropart.\ Phys.}  {\bf 28}, 390 
  [arXiv:astro-ph/0702464].
  
\bibitem{Gelmini:2007jy}
Gelmini G , Kalashev O and Semikoz D V 2007
  {\it JCAP} {\bf 0711}, 002 
  [arXiv:0706.2181 [astro-ph]].
  

    \bibitem{AgasaYakutskLimit}  
  Rubtsov  G I {\it et al.} 2006
 {\it Phys.\ Rev.} D {\bf 73} 063009.
  


\bibitem{Cronin:2007zz}
Abraham J {\it et al.}  [Pierre Auger Collaboration] 2007
 {\it  Science} {\bf 318} 938 
  [arXiv:0711.2256 [astro-ph]].

\bibitem{Abraham:2007si}
  Abraham J {\it et al.}  [Pierre Auger Collaboration] 2008
  {\it Astropart.\ Phys.}  {\bf 29}, 188 
  [Erratum-ibid.\  {\bf 30}, 45 (2008)]
  [arXiv:0712.2843 [astro-ph]].  
  






\bibitem{Abbasi:2008md}
Abbasi R U  {\it et al.} 2008
  {\it Astropart.\ Phys.} {\bf 30} 175 
  [arXiv:0804.0382 [astro-ph]].

  



\end{thebibliography}
\end{document}